\documentclass{article}
\usepackage{spconf,amsmath,graphicx,subfig,footnote,bm,hyperref}

\title{Egocentric Activity Recognition with Multimodal Fisher Vector}
\makeatletter

\name{ Sibo Song$^{\star}$, Ngai-Man Cheung$^{\star}$, Vijay Chandrasekhar$^{\dagger}$, Bappaditya Mandal$^{\dagger}$, Jie Lin$^{\dagger}$}

\address{$^{\star}$Singapore University of Technology and Design, $^{\dagger}$Institute for Infocomm Research, Singapore}

\begin{document}
\topmargin=0mm
\ninept
\maketitle
\begin{abstract}
With the increasing availability of wearable devices, research on egocentric activity recognition has received much attention recently. In this paper, we build a Multimodal Egocentric Activity dataset which includes egocentric videos and sensor data of 20 fine-grained and diverse activity categories.  We present a novel strategy to extract temporal trajectory-like features from sensor data. We propose to apply the Fisher Kernel framework to fuse video and temporal enhanced sensor features. Experiment results show that with careful design of feature extraction and fusion algorithm, sensor data can enhance information-rich video data.  We make publicly available the Multimodal Egocentric Activity dataset to facilitate future research.
\end{abstract}
\begin{keywords}
Multimodal Egocentric Activity dataset, Egocentric activity recognition, Multimodal Fisher Vector
\end{keywords}

\section{Introduction}
Egocentric or first-person activity recognition has attracted a lot of attention. The ever-increasing adoption of wearable devices such as Google Glass, Microsoft SenseCam, Apple Watch and Mi band facilitates low-cost, unobtrusiveness collection of rich egocentric activity data. These devices can monitor activities and gather data anytime from anywhere. The egocentric activity data is helpful in many applications ranging from security, health monitoring, lifestyle analysis to memory rehabilitation for dementia patients.



Research on automatic egocentric activity recognition has been focusing on using two broad categories of data: low-dimensional sensor data and high-dimensional visual data.  Low-dimensional sensor data such as GPS, light, temperature, direction or accelerometer data has been found to be useful for activity recognition \cite{rezaie2015implementation, kwapisz2011activity, guan2007activity, maurer2006activity, khan2014activity}. \cite{kwapisz2011activity} proposes features for egocentric activity recognition computed from cell-phone accelerometer data. They reported over 90\% accuracy for 6 simple activities. \cite{guan2007activity} reported more than 80\% accuracy for 9 activities with sensors located at two legs.
Low-dimensional sensor data can be collected and stored easily, and the computational complexity of the recognition is usually low. 

More recently, there is a lot of interests to perform egocentric activity recognition using high-dimensional visual streams recorded from individuals' wearable cameras.
Compared to low-dimensional sensor data, visual data captures much richer information: scene details, people or objects the individual interacts, for example.
Therefore, several egocentric video datasets and approaches have been proposed to recognize complex activities.
Among them, some previous works focus on extraction of egocentric semantic features like object \cite{pirsiavash2012detecting, fathi2011learning}, gestures \cite{lee2014hand} and object-hand interactions \cite{fathi2011understanding} or discriminative features\cite{mandal20103}. Recently, trajectory-based approach \cite{wang2013dense} has been applied to characterize ego-motion in egocentric videos, and encouraging results have been obtained for activity classification \cite{song2014activity}.

Previous work has investigated egocentric activity recognition using either low-dimensional sensor data or high-dimensional visual data. To the best of our knowledge, there is no previous work to study the potential improvement with using both the sensor and visual data simultaneously.
To address this research gap, we make several novel contributions in this work.  

First, we build and make publicly available a challenging Multimodal Egocentric Activity dataset\footnote{Dataset:\href{http://people.sutd.edu.sg/~1000892/dataset}{http://people.sutd.edu.sg/~1000892/dataset}} that consists of 20 complex and diverse activity categories recorded in both sensor and video data. The sensor and video data are recorded in a synchronized and integrated manner to allow new algorithms to explore them simultaneously.  Second, we propose a novel technique to fuse the sensor and video features using the Fisher Kernel framework.
Fisher kernel combines the strengths of generative and discriminative approaches\cite{jaakkola1999exploiting, perronnin2007fisher}. In this work, we propose a generative model for the sensor and video data. Base on the model, we apply Fisher kernel framework to compute multimodal feature vectors for a discriminative classifier. We refer the determined multimodal feature vectors as Multimodal Fisher Vector (MFV). Third, we perform comprehensive experiments to compare the performance of different approaches: sensor-only, video-only, and the fusion of both.






The rest of this paper is organized as follows. In Section \ref{sec:dataset} we present our Multimodal Egocentric Activity dataset. The methodology is described in Section \ref{sec:method}. Experimental evaluation is presented in Section \ref{sec:eval} and we conclude the work in Section \ref{sec:conclusions}. 

\section{Multimodal Egocentric Activity Dataset}
\label{sec:dataset}

\begin{figure}
\centering
\subfloat[]{\includegraphics[width=0.48\columnwidth]{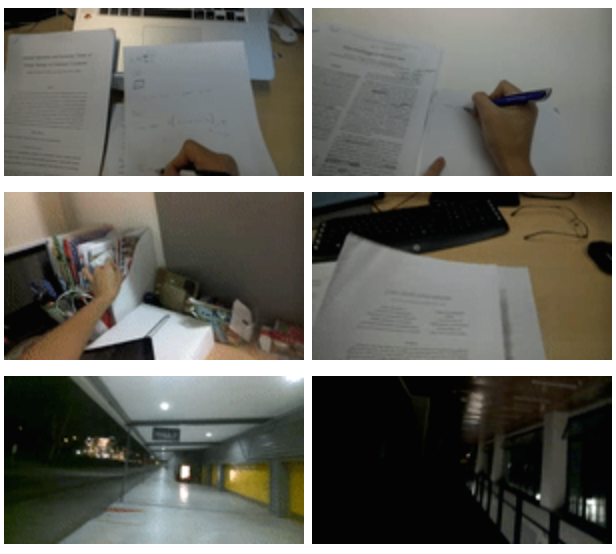} \label{fig:sample}}
\subfloat[]{\includegraphics[width=0.52\columnwidth]{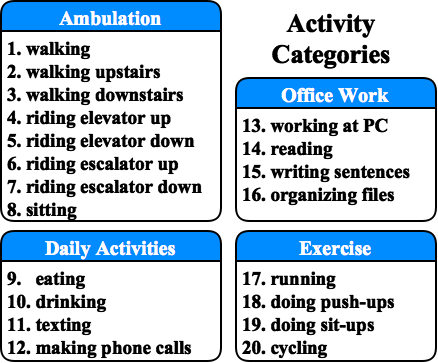} \label{fig:category}} 
\caption{(a) Sample frames from videos (b) Activity categories}
\end{figure}

%
%



To record multimodal egocentric activity data, we created our own application that runs on Google Glass. This application enables us to record egocentric video and sensor data simultaneously in a synchronized manner. The following types of sensor data are supported on the glass: accelerometer, gravity, gyroscope, linear acceleration, magnetic field and rotation vector. These sensors are integrated in the device and they can help capturing accurate head motion of individuals when they are performing different activities.

The Multimodal Egocentric Activity dataset contains 20 distinct life-logging activities performed by different human subjects. Each activity category has 10 sequences. Each clip has exactly a duration of 15 seconds. The activity categories are listed in Fig.\ref{fig:category}. Sample frames of \textit{writing sentences}, \textit{organizing files} and \textit{running} are shown in Fig.\ref{fig:sample}. These categories can be grouped into 4 top level types: \textit{Ambulation}, \textit{Daily Activities}, \textit{Office Work}, \textit{Exercise}. Subjects have been instructed to perform activities in a natural and unscripted way. 
%

The dataset has the following characteristics.
Firstly, it is the first life-logging activity dataset that contains both egocentric video and sensor data which are recorded simultaneously. 
Secondly, the dataset contains considerable variability in scenes and illumination. Videos are recorded both indoor and outdoor, with significant change in the illumination conditions. 
For instance, the walking routes are in various environments, such as residential areas, campus, shopping mall, etc. Thus even though different subjects do the same activity, their background varies. 
Thirdly, we build a taxonomy based on the categories shown in Fig.\ref{fig:category}. All 20 categories can be grouped into 4 top level types. This allows evaluation of new visual analysis algorithms against different levels of life-logging activity granularity. 

\section{Methodology}
\label{sec:method}

In this section, we describe egocentric video and sensor feature extraction. 
Furthermore, we present a simple but effective way to encode temporal information in sensor features. In the end, we propose a Multimodal Fisher Vector approach for combining video and sensor features. 

\subsection{Video trajectory features}

We evaluate state-of-the-art trajectory-based activity recognition on egocentric videos in our own dataset. The dense trajectory approach we used has already been applied to third-person view activity recognition in \cite{wang2013dense}. Dense trajectory approach is applied instead of improved trajectory approach \cite{wang2013action} because dense trajectory could capture more ego-motion information in egocentric videos. 

Dense trajectories are obtained by tracking densely sampled points using optical flow fields. Foreground motions, background motions and head movements are extracted using this approach. Several kinds of descriptors are computed for each trajectory and their characteristics are discussed in \cite{wang2013dense}. Trajectory is a concatenation of normalized displacement vectors. The other descriptors like \textit{MBH} (motion boundary histograms) are computed in the space-time volume aligned with the trajectory. 
These descriptors are commonly used for activity recognition nowadays. Then we applied Fisher kernels on trajectory features as in \cite{song2014activity}, where the trajectory features are represented by means of a Gaussian Mixture Model (GMM). 

\subsection{Temporal enhanced trajectory-like sensor features}

In this section, we establish a novel strategy to convert sensor time-series data into trajectory-like feature. And then temporal order is introduced to enhance sensor feature. 

\begin{figure}
\centering
\includegraphics[width=0.85\columnwidth]{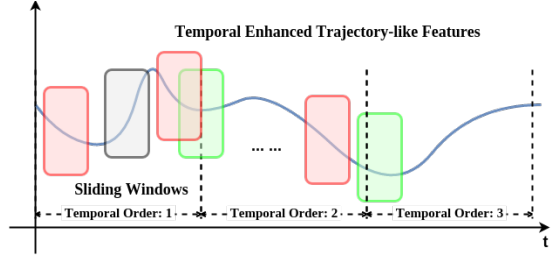} 
\caption{Illustration of temporal enhanced trajectory-like sensor features. (Color of windows indicates similar patterns)}
\label{fig:fisher}
\end{figure}

\textbf{Trajectory-like features} In video analysis, dense trajectories are computed by tracking densely sampled points using optical flow fields. Similarly, we firstly convert time-series data of sensors into trajectory-like data. So a sliding window is required to generate trajectory of sensor data (See Fig.\ref{fig:fisher}). For example, we build a window of 10 samples to extract a trajectory of length equals to 10 and then move the window one frame forward and extract another trajectory. This process could be done until end of the file and thus we could generate many trajectories for sensor data. 

\textbf{Temporal enhanced sensor features} Generally, in image processing, Fisher vector encoding is an orderless approach because it represents an image as an orderless collection of features. While, for sensor data processing, the temporal order of trajectory-like data or patterns could be significantly different in recognizing two activities. However, we do not want timestamp of each trajectory to be included since patterns do not hold exact correspondence in time sequences for one activity. If we simply add timestamp as one more dimensional data, noises could also be added into sensor features. Because essentially the precise timing of a pattern does not matter. While, in some cases, the order of patterns or the stage of pattern occurrence plays an major role in differentiating two distinct activities. 

To overcome the difficulty, we segment the time-series data into several stages to avoid fine-grained windows. And then these segments are indexed starting from $1$ which represents their temporal orders as Fig.\ref{fig:fisher} shows. Then the trajectory-like features inside each part is associated with their normalized temporal order. It turns out to be effective with this quantization strategy for activity recognition using sensor data. 

\subsection{Multimodal Fisher Vector}

In this section, we construct a Multimodal Fisher Vector (MFV) which utilizes the Fisher Kernel framework to combine video trajectory features and temporal enhanced sensor features extracted from a clip. Fisher Kernel has been introduced to combine the benefits of generative and discriminative approaches \cite{jaakkola1999exploiting}.
The idea is to characterize signals with a gradient vector derived from a pdf which models the generating process of the signal.  In our case, the signal is the tuple $\bm{f}=(\bm{x},\bm{s})$, $\bm{x}$ being the video trajectory feature and $\bm{s}$ being the associated temporal enhanced sensor feature.  The generative model of $\bm{f}$ will be discussed later.

Let $p$ be a pdf with a parameter set $\lambda$. Under the framework, one can characterize $N$ features $\bm{F}=\left\{\bm{{f}_{n}},n=1,\cdots,N\right\}$ with the following gradient vector:

\begin{equation}
{\nabla}_{\lambda}\log{p(\bm{F}|\lambda)}
\end{equation}

To illustrate MFV,  we first consider video features codebook generated by a Gaussian Mixture Model (GMM). $\bm{X}=\left\{\bm{{x}_{n}},n=1,\cdots,N\right\}$ denotes the set of trajectory features extracted from videos and $\lambda$ the set of parameters of the GMM. $\lambda=\left\{{\theta}_{i},\bm{{\mu}_{xi}},\bm{{\Sigma}_{xi}},i=1,\cdots,K\right\}$ where ${\theta}_{i}$,$\bm{{\mu}_{xi}}$, and $\bm{{\Sigma}_{xi}}$ denote respectively the weight, mean vector and covariance matrix of Gaussian $i$ and $K$ denotes the number of Gaussian. We assume diagonal $\bm{{\Sigma}_{xi}}$.
The likelihood that the observation $\bm{{x}_{n}}$ was generated by GMM is:
\begin{equation}
p(\bm{{x}_{n}}|\lambda)=\sum_{i=1}^{K}{{\theta}_{i}{p}_{i}(\bm{{x}_{n}})} 
\end{equation}
the weights are subject to the constraint: $\sum{{\theta}_{i}=1}$. The components ${p}_{i}$ are given by:
\begin{equation}
\label{eq:gaussian_1}
p(\bm{x}|\omega=i)=\frac{\exp(-\frac{1}{2}{(\bm{x}-\bm{\bm{{\mu}_{xi}}})}'\bm{{\Sigma}_{xi}}^{-1}{(\bm{x}-\bm{{\mu}_{xi}})})}{{(2\pi)}^{D/2}{\left|\bm{{\Sigma}_{xi}}\right|}^{1/2}} 
\end{equation}
where $D$ is the dimensionality of the video feature vectors and $\left|\bullet\right|$ denotes the determinant operator.

We parameterize ${\theta}_{i}=\exp({\alpha}_{i})/\sum_{j}^{}{\exp({\alpha}_{j})}$, which avoids enforcing explicitly the constraints. With ${q}_{ni}$ to denote the posterior $p(\omega=i|\bm{{x}_{n}})$, or responsibility, and $\bm{{x}_{ni}}$ to denote $\bm{{x}_{n}}-\bm{\bm{{\mu}_{xi}}}$, the gradients of the log-likelihood for a single trajectory feature are,
\begin{equation}
\frac{\partial\log{p(\bm{{x}_{n}})}}{\partial{\alpha}_{i}}={q}_{ni}-{\theta}_{i}
\end{equation}
\begin{equation}
\frac{\partial\log{p(\bm{{x}_{n}})}}{\partial\bm{{\mu}_{xi}}}={q}_{ni}\bm{{\Sigma}_{xi}}^{-1}\bm{{x}_{ni}}
\label{eq:pd1}
\end{equation}
\begin{equation}
\frac{\partial\log{p(\bm{{x}_{n}})}}{\partial\bm{{\Sigma}_{xi}}^{-1}}={q}_{ni}(\bm{{\Sigma}_{xi}}^{}-\bm{{x}_{ni}}^{2})/2
\label{eq:pd2}
\end{equation}
$\bm{{x}_{ni}}^{2}$ denotes element-wise square.
The representation is obtained by averaging these gradients over all $\bm{{x}_{n}}$. 

To generate MFV, we apply a single Gaussian as the generative model for sensor features with the same label $\omega$. To assign $\omega$ to sensor features, we first regard GMM as the basic bag-of-words model by labeling the video feature points according to their largest $p(\omega=i|\bm{{x}_{n}})$. 
It means we use the sensor patterns which has the same starting time with video trajectories in video codebook to build sensor codebook instead of using GMM. The intuition is that the visual trajectories can be associated with sensor trajectory-like features by establishing relationship between two modalities' codebooks. 
In other words, one single Gaussian of sensor trajectory-like features is built based on each Gaussian of video trajectory features when constructing the codebook for sensor features.

Each sensor feature could correspond to multiple video trajectories, because a lot of video trajectories could be generated for each frame. So we use max pooling to assign label $\omega$ to the sensor data. Each vector is then represented as the tuple $\bm{b}=(\omega,\bm{s})$, where $\omega$ is the label and $\bm{s}$ is the sensor feature associated with temporal order. We then define a generative model over the tuple as,
\begin{equation}
p(\bm{b})=p(\omega)p(\bm{s}|\omega)
\end{equation}
\begin{equation}
p(\omega=i)={\theta}_{i}
\end{equation}
\begin{equation}
\label{eq:gaussian_2}
p(\bm{s}|\omega=i)=\frac{\exp(-\frac{1}{2}{(\bm{s}-\bm{{\mu}_{si}})}'\bm{{\Sigma}_{si}}^{-1}{(\bm{s}-\bm{{\mu}_{si}})})}{{(2\pi)}^{(d+1)/2}{\left|\bm{{\Sigma}_{si}}\right|}^{1/2}} 
\end{equation}
It can be shown that the gradients of the log-likelihood of sensor features have similar representation as video trajectory features, i.e., (\ref{eq:pd1}), (\ref{eq:pd2}). We assume that the video feature $\bm{x}$ and sensor feature $\bm{s}$ are independent conditioning on $\omega$. 
We use the following as the generative model for $\bm{f}=(\bm{x},\bm{s})$:
\begin{equation}
p(\bm{f})=\sum_{i}^{}{{\theta}_{i}p(\bm{x}|\omega=i)p(\bm{s}|\omega=i)}
\end{equation}
$p(\bm{x}|\omega=i)$ is defined exactly the same as in (\ref{eq:gaussian_1}) and $p(\bm{s}|\omega=i)$ as in (\ref{eq:gaussian_2}).  The MFV has size $(1+2(D+(d+1)))K=(2D+2d+3)K$ where $D$ is the reduced dimensionality of video trajectory feature using Principal Component Analysis (PCA), and $(d+1)$ the reduced dimensionality of temporal enhanced sensor trajectory-like feature. 

\section{Experimental Evaluation}
\label{sec:eval}

In this section, we first describe our experiment setup. In addition, we study the parameter setting for temporal enhanced trajectory-like sensor feature. Then we evaluate the performance of our MFV method on our dataset. We present and discuss the results for activity recognition task using MFV.

\subsection{Experiment setup}

In our experiment, the dimension of videos is $320\times180$ and frame rate is $10$fps. For sensor data, in total, there are $19$ dimensions of sensor data including accelerometer, gravity, gyroscope, linear acceleration and magnetic field and rotation vector. And in all cases, we collected sensor data in $10$Hz.

The number of Gaussian is set to $K=25$ for video data and we randomly sample a subset of $1\%$ of all samples to estimate GMM for building codebook. The dimensionality of the feature is reduced by half using PCA. Finally, we apply power and $L2$ normalization to MFV as in \cite{perronnin2010improving}. The cost parameter $C=10$ is used for linear Support Vector Machine (SVM) and one-against-rest approach is utilized for multi-class classification. We use \textit{libsvm} library \cite{CC01a} to implement the SVM algorithm.

\subsection{Parameter setting}

\begin{figure}
\centering
\subfloat[]{\includegraphics[width=0.46\columnwidth]{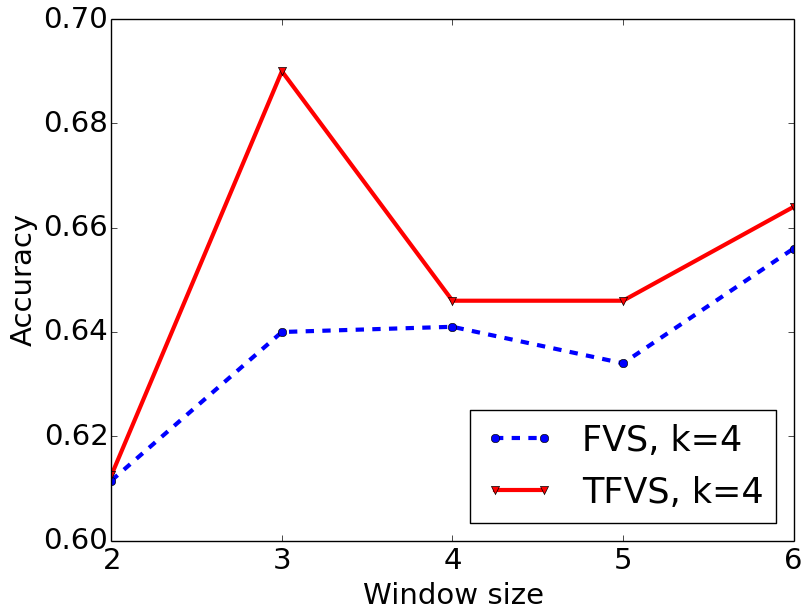} \label{fig:param_w}}
\subfloat[]{\includegraphics[width=0.46\columnwidth]{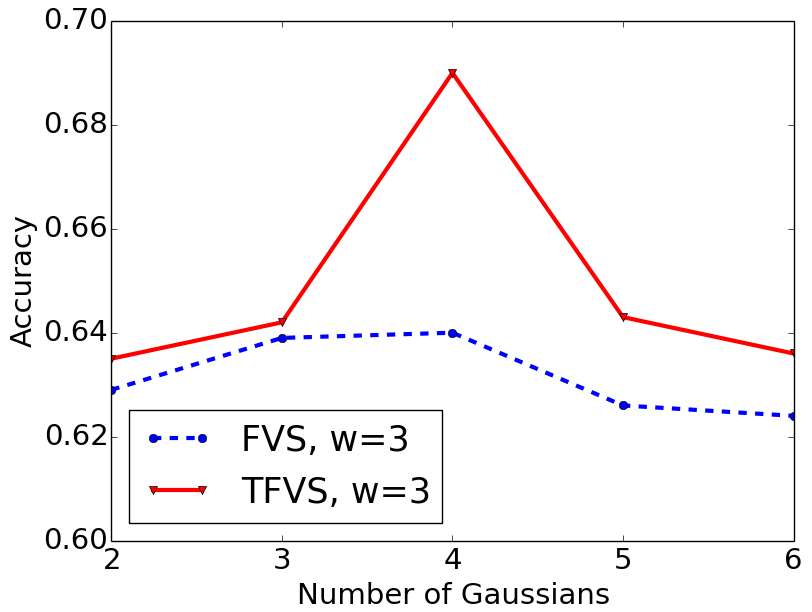} \label{fig:param_k}}
\caption{Comparison of FVS and TFVS with (a) different $w$ (b) different $k$}
\label{fig:param}
\end{figure}

Here, we conduct an experiment to select the window size and number of clusters in GMM process since these parameters are crucial in trajectory-like sensor features generation. 

We choose a moderate number of segments which is $4$ for indexing temporal orders. In Fig.\ref{fig:param} we show that with window size $w=3$ and number of clusters $k=4$ we could obtain the best result. We compare temporal enhanced sensor Fisher vectors (TFVS) and normal sensor Fisher vectors (FVS) with different parameters. It indicates that usually TFVS outperforms FVS with carefully-selected parameters. And then to evaluate MFV, we also choose $w=3$ for sensor trajectory-like feature.

An important observation here is that usually we do not need large $w$ and $k$. The setting $w=3$ is able to provide enough statistics of different patterns. We notice that $k=4$ achieves the best performance for both FVS and TFVS which makes intuitive sense since sensor data has only $19$ dimensions in each recording which are much smaller compared to video features and, therefore, it does not need a large $k$ to encode the information. This is desirable, because the smaller $k$ used, the less storage GMM models take and the faster the computation goes when utilizing TFVS for encoding. 

\subsection{Evaluation of temporal enhanced sensor Fisher Vector}

We evaluate the proposed TFVS on sensor data in our dataset. To provide comprehensive analysis of different approaches, we further compare the results of the proposed approach with FVS and other methods on activity recognition.

Fig.\ref{fig:cm_sensor} provides the confusion matrix on classifying sensor data using TFVS. The indices of activities are shown in Fig.\ref{fig:category}. We notice that TFVS performs well as expected for some low-level activities like \textit{Ambulation} and \textit{Exercise}. While for other high-level activities like \textit{organizing files}, the accuracies are around $50\%$ only. We believe this is due to the fact that sensor data can only capture the head motion and cannot analyze the details of scene or objects. Note that, TFVS outperforms FVS which confirms that the temporal enhanced method is effective. 

\begin{savenotes}
\begin{table}[]
\centering
\caption{Comparison of different methods}
\label{table:compare}
\scalebox{0.8}{
\begin{tabular}{p{1.3cm}|p{1.2cm}|p{1.9cm}|p{1.8cm}|p{1.2cm}}
\hline
     & SVM\footnotemark[2] & Decision Tree\footnote[2]{feature-based approach in \cite{kwapisz2011activity}} & FVS &  TFVS  \\ \hline
Accuracy & 47.75\% & 51.80\%     & 65.60\%      & \textbf{69.00\%}    \\ \hline \hline
     & FVV     & FVV + FVS & FVV + TFVS &  MFV  \\ \hline
Accuracy & 78.44\% & 80.45\%     & 82.20\%      & \textbf{83.71\%}    \\ \hline
\end{tabular}}
\end{table}
\end{savenotes}

\textbf{Comparison with feature-based approach on sensor data} We evaluate a system that uses phone-based accelerometers and feature-based technique from \cite{kwapisz2011activity} to compare with our TFVS. Features are extracted by calculating statistics like average value, standard deviation and designing other features like time between peaks. Then informative features based on raw accelerometer readings are generated. We choose decision trees and SVM to evaluate these features. The results are reported in Table \ref{table:compare}. We can see that our approach outperforms feature-based approach with both SVM and decision tree classifiers. Because these features are mostly global statistics. While usually only some repeated patterns of the signal that can tell what activity it is. Our TFVS makes use of windows to extract the interesting parts. And with temporal orders, it can get more detailed information for these patterns.

\subsection{Evaluation of Multimodal Fisher Vector}

We conduct experiments to confirm superiority of our MFV representation over both FVV and TFVS, as well as a simple concatenation of these two vectors. We also discuss the behavior of the MFV on our dataset, particularly on some categories. 

\begin{figure}
\centering
\subfloat[]{\includegraphics[width=0.34\columnwidth]{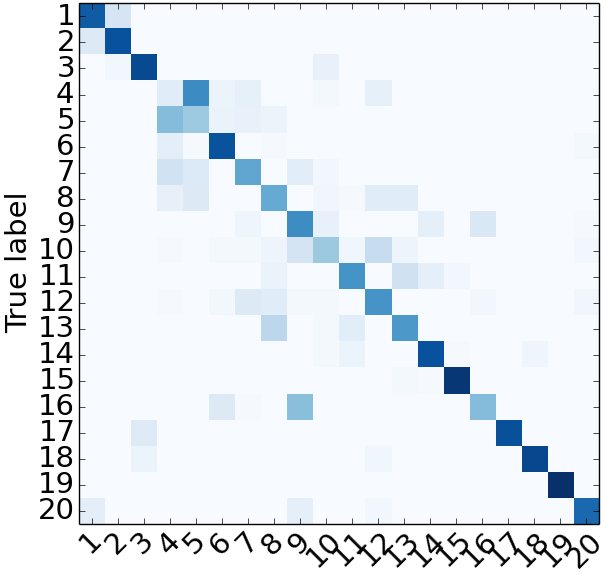} \label{fig:cm_sensor}}
\subfloat[]{\includegraphics[width=0.32\columnwidth]{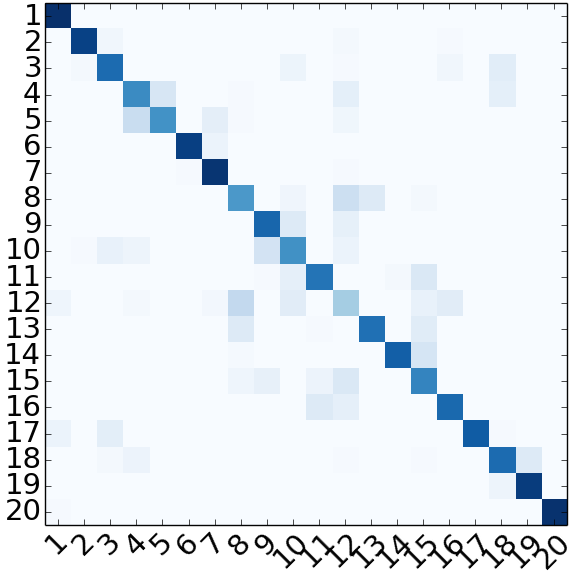} \label{fig:cm_video}}
\subfloat[]{\includegraphics[width=0.32\columnwidth]{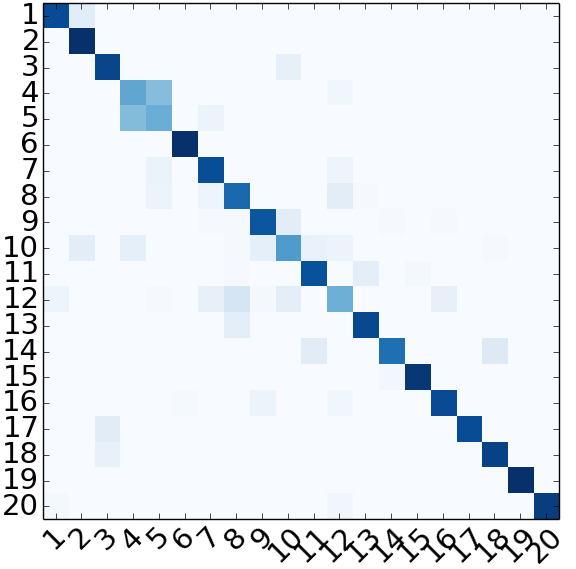} \label{fig:cm_all}}
\caption{Confusion matrices of (a) TFVS (b) FVV (c) MFV }
\end{figure}

The classification accuracy of normal video Fisher vectors (FVV) is reported in Table \ref{table:compare}. By examining the performance of different categories using FVV, we can find that the accuracy of \textit{making phone calls} is $28.57\%$ which is the lowest in Fig.\ref{fig:cm_video}. Videos of \textit{making phone calls} activity are often mis-classified into \textit{sitting} and \textit{drinking}. In fact, for \textit{making phone calls}, it is hard to recognize even for human beings since there is hardly any object appearance in the scene. 

Fig.\ref{fig:cm_all} shows the confusion matrix for recognizing activities using MFV. Actually, it looks similar with confusion matrix for FVV. 
While there are still improvements, because sensors can provide accurate data for slight head motion. By comparing confusion matrices of FVV and MFV, we observe that the largest improvements occur in \textit{making phone calls} ($51.1\%$), \textit{working at PC} ($22.1\%$) and \textit{walking downstairs} ($12.0\%$). For \textit{making phone calls}, it is difficult for visual-based approaches to classify since there is hardly any object appearance in subjects' view. And with temporal enhanced sensor data, we can find that there does exist some patterns for head movement while making phone calls which is a very interesting observation. And for \textit{working at PC}, we believe that video trajectory approach is confused by some movements on the monitors. 
These results again confirm that our MFV approach performed better than FVV and it is more accurate and stable.

From Table \ref{table:compare}, we are able to observe that our MFV performs superior to concatenation of FVV and FVS and even concatenation of FVV and TFVS which has an accuracy of $82.2\%$. The results suggest that MFV is a better representation to combine video and sensor data. Intuitively, the single Gaussian model takes advantage of detailed sensor statistics in each cluster of GMM built for video trajectory features. Furthermore, there is a possibility of extending single Gaussian model to multiple Gaussian models in order to improve the accuracy. 

\subsection{Computational cost}

From Table \ref{table:compare}, we can see that by encoding temporal enhanced sensor features, we could improve the accuracy from $78.4\%$ to $83.7\%$. And this is helpful especially when computing Fisher vector on sensor data is extremely fast since its volume is quite small compared to video data. In practical, generating trajectory features and calculating FV on all videos in the dataset takes about 3 hours, while it only takes less than 1 minute for all sensor data. And with only TFVS we could obtain an accuracy of $69.0\%$ even for 20 categories, which already satisfies the needs of some applications. 

Therefore, there is no single perfect answer. Depending on particular application and requirement, we could choose different solutions. MFV usually provides the best performance, while it is expensive for processing. In some cases, TFVS can achieve satisfactory performance with limited computational ability.

\section{Conclusions}
\label{sec:conclusions}

We built a challenging Multimodal Egocentric Activity dataset which is recorded with videos and sensor data. The dataset contains a variety of scenes and personal styles, capturing the diversity and complexity of life-logging activities. We performed a detailed evaluation of state-of-the-art trajectory-based technique on videos and achieve accuracy around $78\%$. We propose a novel technique to incorporate temporal information into trajectory-like sensor data. Our TFVS outperforms the basic feature-based approach significantly and also the FVS. More importantly, we propose to apply Fisher Kernel framework to fuse sensor and video data.  
We improve the performance from $78.4\%$ to $83.7\%$ by utilizing MFV for egocentric activity recognition at a low cost.

\vfill
\pagebreak

\bibliographystyle{IEEEbib}
\bibliography{egbib}

\begin{thebibliography}{10}

\bibitem{rezaie2015implementation}
Hamed Rezaie and Mona Ghassemian,
\newblock ``Implementation study of wearable sensors for activity recognition
  systems,''
\newblock {\em Healthcare Technology Letters}, vol. 2, no. 4, pp. 95--100,
  2015.

\bibitem{kwapisz2011activity}
Jennifer~R Kwapisz, Gary~M Weiss, and Samuel~A Moore,
\newblock ``Activity recognition using cell phone accelerometers,''
\newblock {\em ACM SigKDD Explorations Newsletter}, vol. 12, no. 2, pp. 74--82,
  2011.

\bibitem{guan2007activity}
Donghai Guan, Weiwei Yuan, Young-Koo Lee, Andrey Gavrilov, and Sungyoung Lee,
\newblock ``Activity recognition based on semi-supervised learning,''
\newblock in {\em Embedded and Real-Time Computing Systems and Applications,
  2007. RTCSA 2007. 13th IEEE International Conference on}. IEEE, 2007, pp.
  469--475.

\bibitem{maurer2006activity}
Uwe Maurer, Asim Smailagic, Daniel~P Siewiorek, and Michael Deisher,
\newblock ``Activity recognition and monitoring using multiple sensors on
  different body positions,''
\newblock in {\em Wearable and Implantable Body Sensor Networks, 2006. BSN
  2006. International Workshop on}. IEEE, 2006, pp. 4--pp.

\bibitem{khan2014activity}
Adil~Mehmood Khan, Ali Tufail, Asad~Masood Khattak, and Teemu~H Laine,
\newblock ``Activity recognition on smartphones via sensor-fusion and kda-based
  svms,''
\newblock {\em International Journal of Distributed Sensor Networks}, vol.
  2014, 2014.

\bibitem{pirsiavash2012detecting}
Hamed Pirsiavash and Deva Ramanan,
\newblock ``Detecting activities of daily living in first-person camera
  views,''
\newblock in {\em Computer Vision and Pattern Recognition (CVPR), 2012 IEEE
  Conference on}. IEEE, 2012, pp. 2847--2854.

\bibitem{fathi2011learning}
Alireza Fathi, Xiaofeng Ren, and James~M Rehg,
\newblock ``Learning to recognize objects in egocentric activities,''
\newblock in {\em Computer Vision and Pattern Recognition (CVPR), 2011 IEEE
  Conference On}. IEEE, 2011, pp. 3281--3288.

\bibitem{lee2014hand}
Sang-Rim Lee, Sven Bambach, David~J Crandall, John~M Franchak, and Chen Yu,
\newblock ``This hand is my hand: A probabilistic approach to hand
  disambiguation in egocentric video,''
\newblock in {\em Computer Vision and Pattern Recognition Workshops (CVPRW),
  2014 IEEE Conference on}. IEEE, 2014, pp. 557--564.

\bibitem{fathi2011understanding}
Alireza Fathi, Ali Farhadi, and James~M Rehg,
\newblock ``Understanding egocentric activities,''
\newblock in {\em Computer Vision (ICCV), 2011 IEEE International Conference
  on}. IEEE, 2011, pp. 407--414.

\bibitem{mandal20103}
Bappaditya Mandal and How-Lung Eng,
\newblock ``3-parameter based eigenfeature regularization for human activity
  recognition,''
\newblock in {\em Acoustics Speech and Signal Processing (ICASSP), 2010 IEEE
  International Conference on}. IEEE, 2010, pp. 954--957.

\bibitem{wang2013dense}
Heng Wang, Alexander Kl{\"a}ser, Cordelia Schmid, and Cheng-Lin Liu,
\newblock ``Dense trajectories and motion boundary descriptors for action
  recognition,''
\newblock {\em International Journal of Computer Vision}, vol. 103, no. 1, pp.
  60--79, 2013.

\bibitem{song2014activity}
Sibo Song, Vijay Chandrasekhar, Ngai-Man Cheung, Sanath Narayan, Liyuan Li, and
  Joo-Hwee Lim,
\newblock ``Activity recognition in egocentric life-logging videos,''
\newblock in {\em Computer Vision-ACCV 2014 Workshops}. Springer, 2014, pp.
  445--458.

\bibitem{jaakkola1999exploiting}
Tommi Jaakkola, David Haussler, et~al.,
\newblock ``Exploiting generative models in discriminative classifiers,''
\newblock {\em Advances in neural information processing systems}, pp.
  487--493, 1999.

\bibitem{perronnin2007fisher}
Florent Perronnin and Christopher Dance,
\newblock ``Fisher kernels on visual vocabularies for image categorization,''
\newblock in {\em Computer Vision and Pattern Recognition, 2007. CVPR'07. IEEE
  Conference on}. IEEE, 2007, pp. 1--8.

\bibitem{wang2013action}
Heng Wang and Cordelia Schmid,
\newblock ``Action recognition with improved trajectories,''
\newblock in {\em Computer Vision (ICCV), 2013 IEEE International Conference
  on}. IEEE, 2013, pp. 3551--3558.

\bibitem{perronnin2010improving}
Florent Perronnin, Jorge S{\'a}nchez, and Thomas Mensink,
\newblock ``Improving the fisher kernel for large-scale image classification,''
\newblock in {\em Computer Vision--ECCV 2010}, pp. 143--156. Springer, 2010.

\bibitem{CC01a}
Chih-Chung Chang and Chih-Jen Lin,
\newblock ``{LIBSVM}: A library for support vector machines,''
\newblock {\em ACM Transactions on Intelligent Systems and Technology}, vol. 2,
  pp. 27:1--27:27, 2011.

\end{thebibliography}

\end{document}